Collaborative Visualization

# AstroSim: Collaborative Visualization of an Astrophysics Simulation in Second Life

Arturo Nakasone, Helmut Prendinger, Simon Holland, and Kenichi Miura ■ *National Institute of Informatics, Japan*

Piet Hut ■ *Institute for Advanced Study*

Jun Makino ■ *National Astronomical Observatory of Japan*

Natural-sciences researchers have created innovative, powerful computational tools to analyze the large amount of data their experiments generate. Commonly, these tools use visualization techniques that help scientists better understand their data and obtain a complete picture of the situation. Through visualizations, scientists can more easily generate and test hypotheses, which otherwise would be time consuming or impractical. Because researchers and experts are often geographically distributed and sparse, demand for collaboration in scientific data visualization is growing. So, scientists need intuitive collaborative environments for natural interaction and data sharing.

Several initiatives aim to advance scientific collaboration by providing powerful networking infrastructure and tools. The Cyber Science Infrastructure, for example, is a comprehensive framework the National Institute of Informatics and other Japanese institutions implemented in 2004 to create an information-technology-based environment for advancing scientific research, collaboration, and education.[1] The US National Science Foundation (NSF) launched the Cyber-Enabled Discovery and Innovation initiative in 2008 to explore new approaches at the intersection of computation and the life sciences.[2] These initiatives, among others, demonstrate collaboration's importance in advancing scientific R&D of better high-end tools that let researchers visualize, share, and discuss their results.

Our research intends to contribute to those international efforts by providing an environment for synchronous collaborative visualization for astrophysics, using standard equipment (a networked computer). Most tools for astronomers are standalone, highly customized solutions. However, given the relatively small number of experts in each astronomy field, a platform for live collaboration is essential. (For a brief look at the main approaches to collaborative visualization, see the "Trends in Collaborative Visualization Systems" sidebar.)

For a visualization platform, we use the popular 3D multiuser online environment of Second Life (www.secondlife.com). Users, graphically represented as avatars, can communicate and collaborate in real time to coexperience the world of astrophysics. Specifically, our AstroSim system supports important activities and features for research and education, such as data zooming, playing back stellar simulations, and color-coding properties or stars. AstroSim's salient collaborative features include the ability to manipulate stars' visual properties for annotation and to point and refer to stars.

Developed in the Second Life 3D online multiuser environment, AstroSim (astrophysics simulation) provides synchronous collaborative visualization for astronomers. Users can play, halt, and rewind simulations and annotate stars interactively to track individual stars and gain a better understanding of stellar dynamics and astrophysics phenomena.

# Trends in Collaborative Visualization Systems

Although collaboration tools all aim to provide users with connectivity and knowledge-sharing functionality, approaches to achieving this goal differ greatly. Two main factors help classify each approach's contribution. The first is the time frame in which collaboration takes place—that is, whether all the participants perform the collaborative task simultaneously or at different times. The second is the participants' location—that is, whether they're geographically colocated or distributed. This classification method is called Applegate's place-time matrix.[1] Figure A shows the matrix, along with classification examples of common collaborative solutions.

### Same Time, Same Place
The classic approach to collaboration falls into the *same-time, same-place* category. This approach includes classrooms and conference rooms specifically furnished to support collaborative tasks among the participants. In astrophysics simulation, the amount of generated data is generally huge, making visualization tools especially important. Significant technical improvements, ranging from playback enhancements to better particle-display algorithms[2] and scaling methods,[3] help users visualize the intricate relationship between the variables in the simulation computation.

This approach's obvious limitation is that all participants are unlikely to be physically present in the same location at the same time. This constraint is especially true for scientific research, given the globalization of research activities and researchers' dissimilar schedules and traveling costs.

### Different Time, Different Place
Communication-engineering improvements have given rise to more-sophisticated collaborative solutions. In such solutions, information exchange occurs through communication networks, thereby overcoming the difficulties of sharing knowledge among geographically distributed participants. Early on, information exchange occurred mainly through faxes and email. As the Web's popularity grew, so did powerful visualization tools. Added interactivity and processing capabilities let people exchange and display significant amounts of scientific research data.[4] Such *different-time, different-place* collaboration assumes that participants can perform collaborative tasks asynchronously.

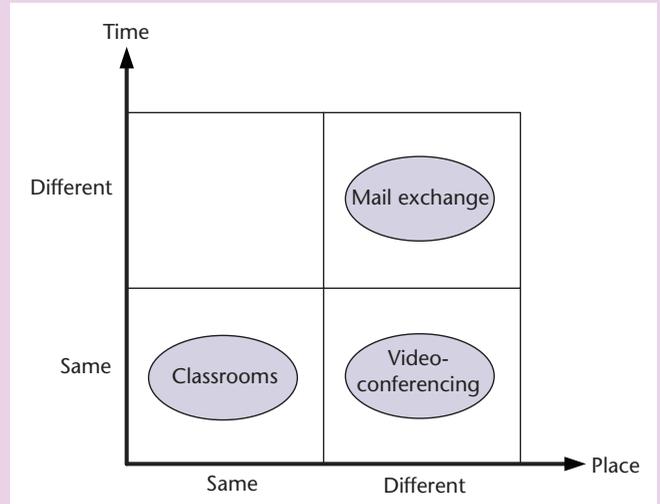

Figure A. Applegate's place-time matrix.[1] Place and time help classify collaboration approaches, such as the common solutions in this matrix.

### Same Time, Different Place
Collaboration in science and engineering requires the capability for all participants, regardless of physical location, to share the same context and experience. Videoconferencing

## Virtual Worlds as Visualization Environments

Thanks to advancements in 3D-graphics processing, virtual worlds are among the most promising visualization tools for collaborative research. Most important, they can enhance collaborative techniques through multimodal character interactions, and they can enrich the visualization experience beyond the constraints of real-world interactive visualizations. The NSF Cyber-Enabled Discovery and Innovation initiative emphasizes virtual environments as important devices to enhance discovery, innovation, and learning.[2]

For implementing collaborative systems, virtual worlds appear more suitable than 3D game engines for two reasons.[3] First, users can experience a seamless persistent world without having to comply with any fixed objectives or rules of a particular game. Second, they can generate, store, and own the content they create.

With millions of active users worldwide, most virtual worlds are open to everyone and serve as stages for the development of social networking, commerce, education, entertainment, and, more recently, scientific research. The most popular of these virtual worlds is Linden Labs' Second Life, an environment in which many people, corporations, and academic institutions have created a virtual presence.

The main objective for implementing scientific visualization in virtual worlds is to let geographically dispersed researchers share virtual space for experiencing, exploring, interacting with, and discussing their data.[4] To fulfill this objective, any application supporting synchronous collaborative data visualization in virtual worlds must is one solution in the *same-time, different-place* category,[5] but it requires a special (and expensive) communication infrastructure. Advances in network processing speed make possible such timely transmission of multimedia information, namely audio and video, over the Internet.

However, other areas of multimedia information processing, such as video compression, still require considerable improvement to provide participants with a smooth multimedia experience. Although advanced techniques, including video sprite generation and partial 3D rendering, have successfully reduced data transmission, content synchronization remains an unsolved issue. Finally, hardware and software components are readily available, but the collaboration experience is limited to communication and presentation.

Beyond exchanging insights and data visualizations, scientists often want to interact with data and experimental settings as they would in a laboratory to clarify any misunderstandings about simulations. Virtual worlds, such as Second Life, and 3D online computer games, such as World of Warcraft, are at the forefront of interaction and collaboration.[5] These platforms provide users interesting and unique ways to interact with other avatars and the worlds they've created. In most cases, interactivity is a user-centered experience in the form of object manipulation, but it can also be an omnipresent interactive scheme among the objects themselves. Using physics engines in simulation games, for example, revolutionized the capacity to provide users with a better, richer experience. This experience improved with enhanced key features in the avatar-interaction paradigm, specifically object contact and natural motion.

Our Second Life-based system falls into the same-time, different-place category. Unlike the asynchronous collaborative visualization the Many Eyes system proposed,[6] our approach supports synchronous collaboration. Although Paul Bourke has suggested using Second Life for synchronous collaborative visualization,[7] a generic, scalable approach to collaborative visualization is still missing. To address this need, we developed AstroSim and the Environment Markup Language 3D, a scripting language for object generation and manipulation in virtual worlds.

- process the input of raw or preprocessed data and automatically turn them into visual entities;
- efficiently use graphic algorithms' ability to handle numerous primitives, such as Second Life's eight primitive types: box, cylinder, prism, sphere, torus, tube, ring, and sculpted;
- provide the ability to map colors, texture, and opacity to the created objects to achieve appealing visualizations that support easy comprehension of phenomena;
- enable intuitive navigation of the visualization with virtual collaborators (avatars), and permit forms of communication such as chat, voice, and deictic gesture;
- support editing visualization data in either private or shared mode; and
- support use of multiple hardware and software platforms.

Taking into consideration these technical and nontechnical issues, we propose that an ideal collaborative environment for astrophysics research should do four things. First, it should provide an immersive experience and natural interactivity with the environment. Second, it should offer online discussion and collaborative visualization. Third, it should stream and handle large amounts of data. Finally, it should be easy to use and offer intuitive navigational functions.

## AstroSim

AstroSim is a joint research project between the National Institute of Informatics in Tokyo, the National Astronomical Observatory of Japan, and the Institute for Advanced Study in Princeton, New Jersey. The primary objective is to develop a Second Life-based application in which

## Globular Star Cluster Evolution

Globular star clusters typically contain roughly 1 million stars. They're relatively isolated, describing wide orbits around their parent galaxies, with orbital periods on the order of 100 million years (see Figure B). In contrast, a galaxy can easily contain 100 billion stars. Star clusters in the disk of a spiral galaxy, such as our own Milky Way, generally contain far fewer stars, on the order of a few hundred or sometimes a few thousand.

This difference in number is important. Each quantitative increase by a factor of a thousand or more implies a significant qualitative difference. Small star-forming regions in our galaxy tend to evaporate on a time scale that's short compared with our galaxy's age, which is close to the universe's age. Globular clusters, in contrast, live far longer. The explanation for this longevity concerns the *relaxation effect*, a process in which stars exchange energy through close encounters. This process is analogous to the way energy is distributed in a gas, such as the air in a room: close encounters between molecules cause the thermal energy to spread equally through the room, mixing the gas and wiping out the memory of the initial conditions. These relaxation effects become less important with greater numbers of stars, for which global forces dominate over local forces.

Our galaxy as a whole, with a far greater number of stars than globular clusters contain, has a relaxation time scale that's far longer than the universe's age. This is why the galaxy can show such pretty spiral features: randomizing individual stars' motions on a relaxation time scale would have wiped out these spirals. Globular clusters, with relaxation time scales of only a fraction of a billion years, are dynamically old enough to have forgotten their initial conditions. This is what makes them round (globular), the most generic shape they can assume after thoroughly randomizing their stars' orbits.

Because globular clusters are relatively isolated from the bulk of their parent galaxies and have a relaxation time a bit shorter than their current age, they're ideal laboratories for stellar dynamics. The pilot calculations we report in the main article contain only 1,000 stars, but we aim to extend the number of stars by a factor of several hundred into the realm of globular clusters. For more background, see Douglas Heggie and Piet Hut's research.[1]

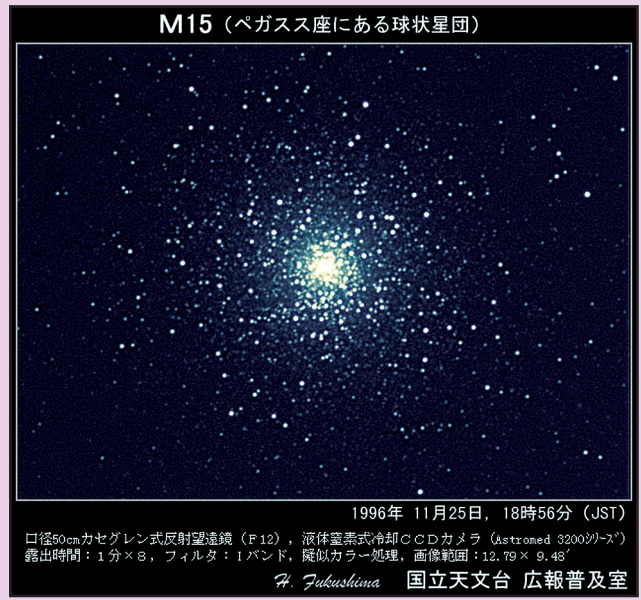

Figure B. Globular star clusters. Clusters such as this M15 Globular Cluster in Pegasus typically contain roughly 1 million stars. (source: National Astronomical Observatory of Japan; used with permission.)

**Reference**
1. D.C. Heggie and P. Hut, *The Gravitational Million-Body Problem*, Cambridge Univ. Press, 2003.

---

researchers can visualize astrophysics phenomena involving simulation of stars' kinematics for research and outreach purposes. Specifically, the researchers use AstroSim for two scenarios of collaborative visualization and data manipulation. This first involves displaying a scaled-down version of a globular star-cluster evolution simulation for educational and presentational purposes (see the "Globular Star Cluster Evolution" sidebar). The second involves displaying an eight-star interaction simulation for research analysis and coexperience.

### System Architecture

Our system comprises two main modules: the *data-processing Web service* and the *visualization-processing Web service* (see Figure 1).

**The data-processing Web service.** After this module receives user requests for a simulation, its control component fulfills them. This component mainly provides an interface for users to interact with the module's functionality. The simulation data originating from the astrophysics simulation server comprises text files containing information about stars' attributes. Each file represents a snapshot in time, in the form of a frame that stores each simulated star's absolute position and its physical properties such as mass, temperature, and velocity. When users request a new simulation, the astrophysics-data-parser component reads, parses, and stores these files. Because not all data files have the same structure, the system uses an external XML-based format specification file to identify the type of data they include.

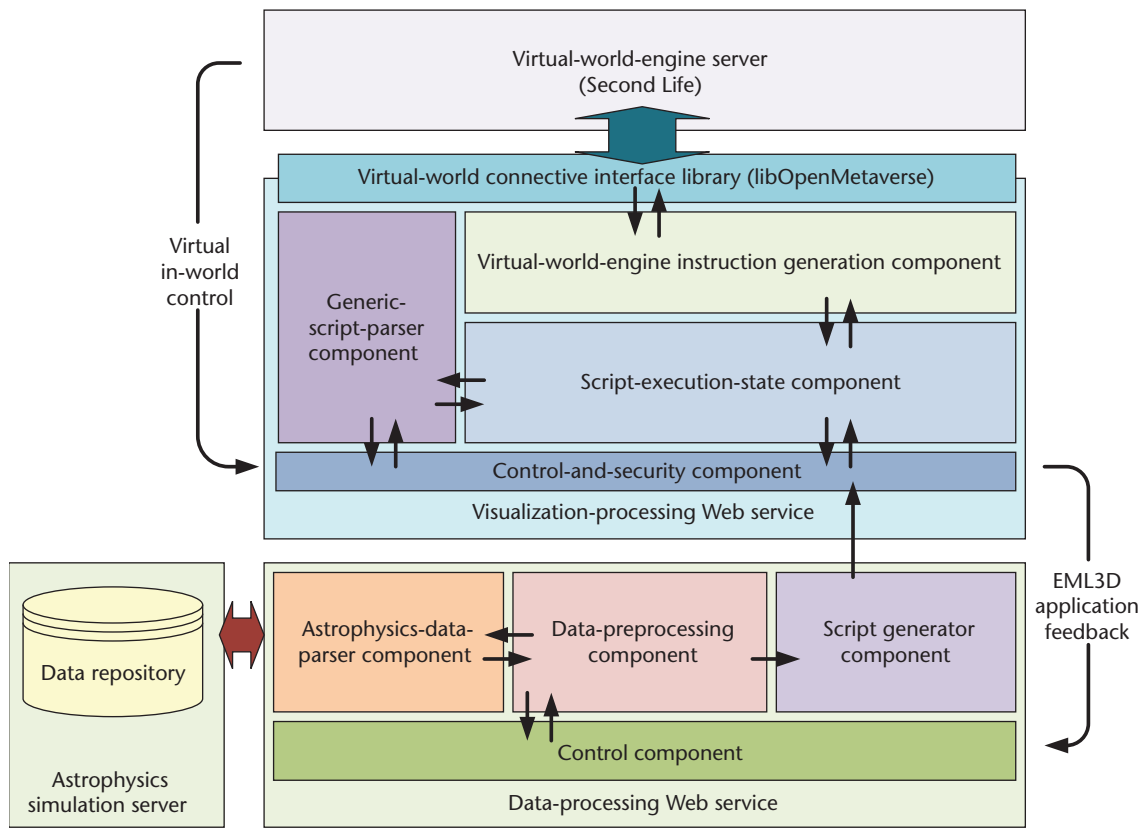

Figure 1. The AstroSim system architecture. The data-processing Web service parses and preprocesses the astrophysics data and sends it as XML-based script streams to the visualization-processing Web service. The latter Web service converts streams to low-level commands to be sent to Second Life through a third-party library called libOpenMetaverse. EML3D stands for Environment Markup Language 3D.

In some cases, the generated frames in the text files present the stars' position values rather sparsely, which can cause jerkiness during their visualization. So, the data-preprocessing component performs a position value interpolation to add smoothness to the frame playback. Next, the stored data moves to the script generator component, which generates a high-level XML-based script stream for each stored frame. These streams travel to the visualization-processing Web service, which parses them and generates the specific Second Life commands to display the objects in the virtual world. Figure 2 shows a raw-data-file sample, which contains position, velocity, temperature, and mass information for a particular set of stars, along with its corresponding script stream.

**The visualization-processing Web service.** To display the stars' physical properties and motion simulation data, the visualization-processing Web service handles the connection between Second Life and AstroSim. This connection is EML3D (Environment Markup Language 3D), a high-level XML-based scripting-language specification that lets users interactively manage their virtual-world environment by creating and manipulating objects through a simple set of instructions. EML3D aims mainly to provide an object-manipulation solution that's independent from a particular virtual world and can process data from external applications such as AstroSim.

Once the control-and-security component receives the script stream from the data-processing Web service, the generic-script-parser component parses and verifies it. If the script has no syntactic or semantic errors, the system stores the instructions to be executed in the script-execution-state component. When the system executes each instruction, the instruction passes to the virtual-world-engine instruction generation component, which generates the necessary low-level commands that pass to Second Life through the virtual-world connective interface library.

AstroSim uses these EML3D capabilities:

- defining several *creator bots* (computer-controlled agents that perform all object manipulation activities in the virtual world) in the script to help reduce the star creation and manipulation load;
- defining the execution of instructions in parallel, thereby reducing script-processing time;
- adding *perceptor bots*, agents that let EML3D process events originating in the virtual world—a feature particularly useful for interaction between avatars and stars;
- creating objects in batch mode, enabling AstroSim to create thousands of objects in a short period; and
- manipulating object properties such as color, texture, light, size, scale, position, and rotation.

Figure 2. A raw data file with its corresponding script stream. At the top left, the raw data file contains information on stars' positions and velocities. On the right, the corresponding script stream describes high-level instructions for star creation and motion. The top script reflects the stars' creation and position; the bottom script shows the coloring process, which is based on temperature values and a scheme defined by the XML-based configuration file at the bottom left.

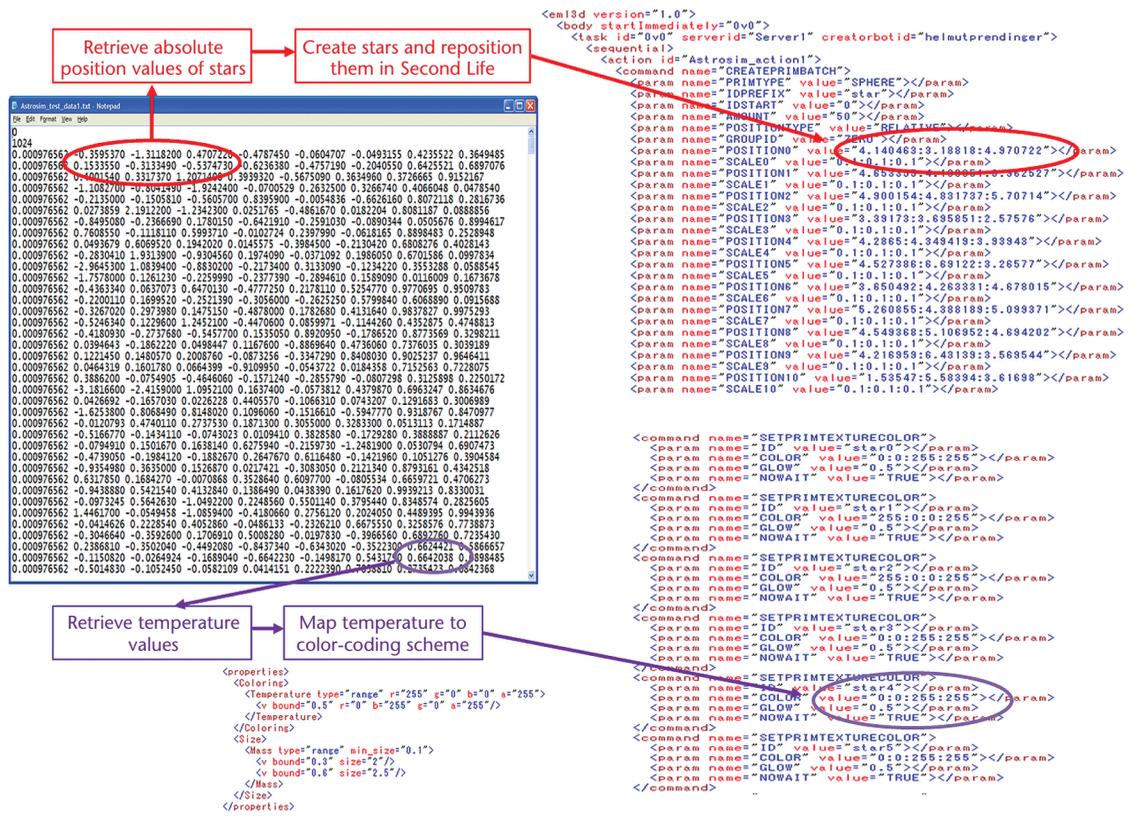

This functionality lets AstroSim process its simulation data conveniently and efficiently by isolating the simulation playback from the technically problematic visualization issues.

### Functional Capabilities

AstroSim fully exploits Second Life's graphical capabilities and lets users flexibly manipulate simulation data. Currently, AstroSim offers the following three main functionalities.

**Automatic star object creation and color-coding on the basis of physical properties.** Stars in our simulation are visualized in the virtual world's open sky (see Figure 3). This depiction creates a dome effect on the avatar, which, in turn, enhances its immersion into the simulation. In astrophysics simulation environments, color-coding schemes help map stars' physical characteristics, such as velocity, mass, and size. AstroSim provides users with a flexible way to define these color schemes. They can define mapping between characteristics and colors in an XML-based external configuration file.

**Simulation playback.** This functionality gives users control over the simulation visualization's execution through simple chat commands. Current functionality lets users play, pause, and rewind the simulation. However, these operations are restricted to dedicated avatars to avoid issuing conflicting commands, such as play and rewind at the same time.

More-advanced functions include changing the playback speed and adjusting the playback's step size. Because the simulation's standard execution can be too fast to analyze, we added a simple delay to let users control the simulation visualization speed. In addition, some simulations can take too much time to complete. By showing only the specific frames defined by the step-size parameter, we let users shorten the simulation time. This feature is useful only when a user wants a general impression of the simulation data, rather than the stars' detailed motion.

**Star-cluster zooming.** One of AstroSim's most interesting features is the ability to provide a better view by zooming in on or out of the data set. Users can obtain a more detailed view of the stars' interaction, which otherwise might be overlooked or be occluded by the cluster. We perform zooming by simply adjusting the scale in which the stars are positioned (see Figure 4). Because the simulation plays in Second Life's open sky, the main restriction for zooming is the Second Life's boundary limit, or *island*. An island is a square land region in the 3D space with 256 meters on each side. This zooming function gives the users considerable flexibility for detailed observation.

However, having the stars expanded throughout the whole island could be problematic not only for visualization—the viewer wouldn't be able to show all the data—but also for manipulation—stars to be manipulated might fall outside the island's boundaries. For our experimental settings, AstroSim displays stars only within set bounds of space inside 10 × 10 × 10 cubic meters (but users can modify this parameter). Anything outside this volume doesn't display. When a star reenters the space set by the boundary limits, it reappears.

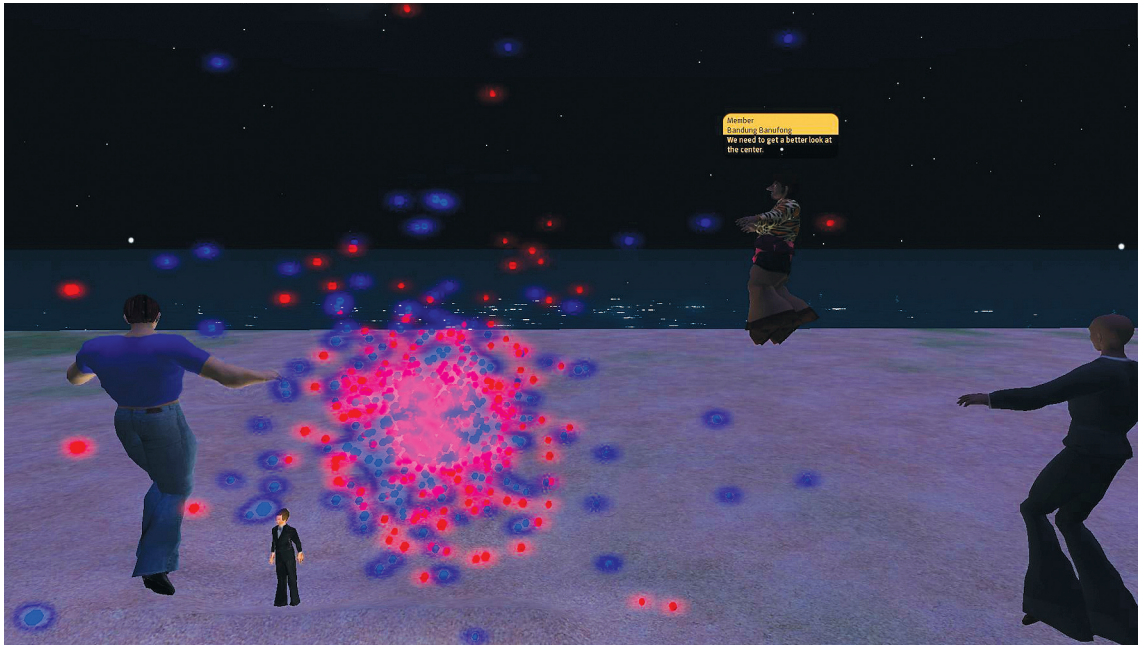

Figure 3. A globular-star-cluster visualization in Second Life. Users can communicate with other avatars by using the Second Life client interface's chat channel. Here, a user represented as the avatar Bandung Banufong types the instant message "We need to get a better look at the center" into Second Life's chat channel.

## Collaborative Capabilities

From the beginning, Second Life's creators built the technology to support synchronous communication and collaboration, because those are key aspects for any real-time social-networking application. This support comprises features such as textual chat, voice transmission, and gesturing. In AstroSim, we used these well-tested features, along with Second Life's capacity for real-time distributed-object manipulation, to enhance our visualization solution's collaborative aspects.

Users can modify star properties, such as scale, color, description, and position, to either track individual stars or better view the observed phenomenon. Because the system generates stars without any Second Life-induced constraints regarding permissions or property modifications, simulation participants can manipulate them (see Figure 5). Manipulating virtual stars in real time can empower researchers with the capability to perform live collaborative visualizations. We also explored other schemes for annotation, such as floating labels, but they require technical expertise, such as in-world scripting. We hope to include this feature in a future version of AstroSim.

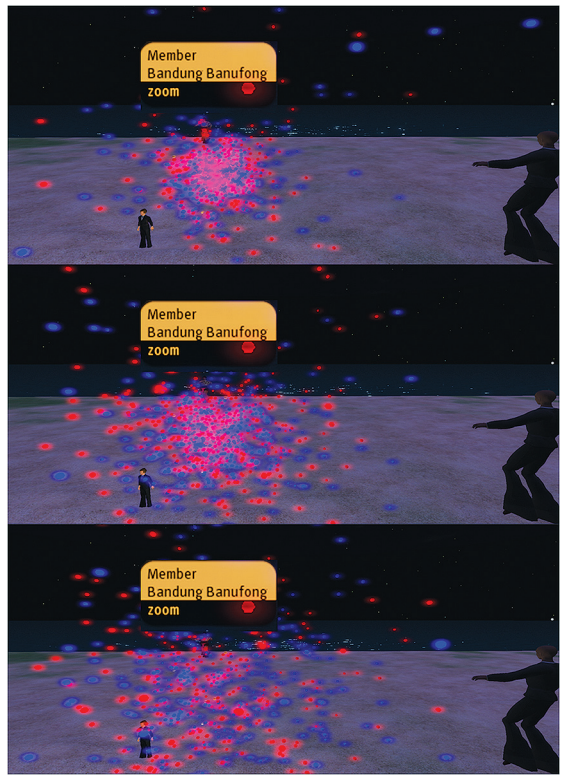

Figure 4. Data zooming. AstroSim interprets words such as "zoom" as commands rather than instant messages. For a better view of the globular star cluster's core, Bandung Banufong types the zoom command. Upon each additional zoom command (top to bottom), AstroSim automatically increments the star display's scale.

AstroSim takes advantage of four other collaborative tools the Second Life client software provides. *Communication logging* lets users record any text-based conversation and use it for analysis, documentation, and discussion to share with colleagues who didn't participate in the simulation. *Gesturing* lets users point to stars to note a particular feature in the simulation data. Combining this tool with star manipulation gives users a powerful way to annotate content and refer to it. To employ a *shared camera view* for comparative analysis, users simply position their avatars next to one another. The final tool is *simulation persistence*; because of Second Life's persistent nature, anyone can rerun a simulation at any time.

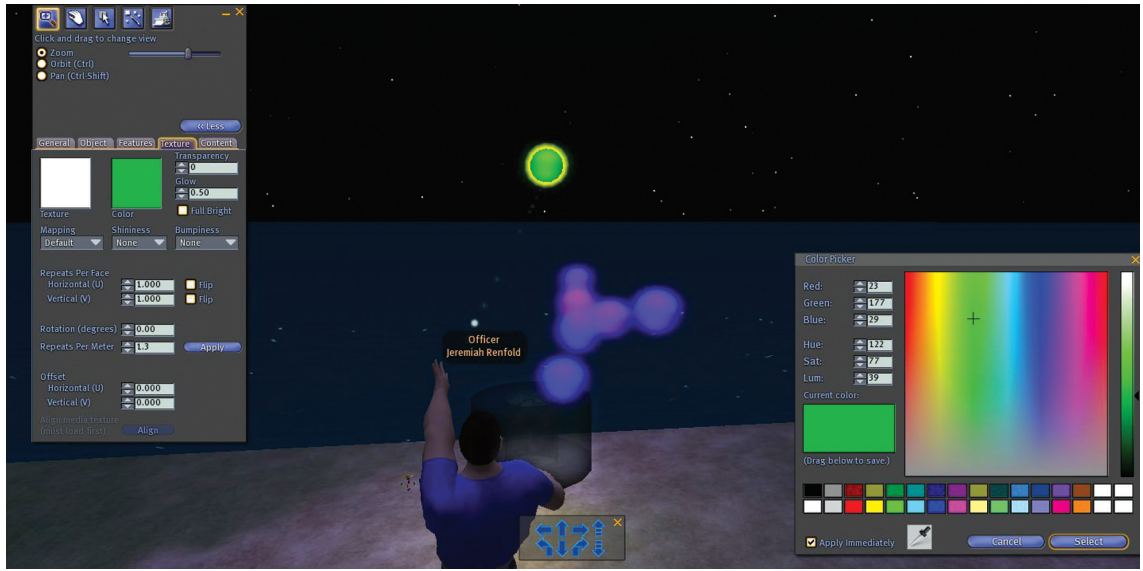

Figure 5. Avatars can manipulate stars to track individual stars or improve a phenomenon's visibility. Users can select stars with a simple right click (provided by the Second Life interface). They can then place the selected stars anywhere in the simulator or change the stars' visual appearance. In the figure, the avatar has colored star STAR0 green to track its movement.

## User Experiences

The most complex behavior of stars in a globular star cluster occurs in the cluster's core. That's where most of the energy exchange takes place, between the external kinetic energy of motion and internal binding energy of double stars and triple stars. To study interactions among single, double, and triple stars in great detail, we only need to model a small part of a larger cluster's core.

This kind of research is of great interest in understanding the formation of double stars, which play an important role in stellar evolution. Modeling the nuclear reactions and other physical processes of single stars is straightforward compared with modeling double stars because double stars can exchange mass and perturb each other's evolution. So, without closely analyzing these processes, obtaining a detailed understanding of globular star cluster evolution is difficult.

We carried out three related collaborative sessions in AstroSim. First, Piet Hut and Jun Makino, astronomers who are experts in stellar dynamics, studied a globular star cluster's dynamics. Second, Hut showed and discussed results with fellow researchers specialized in other astronomy fields. Finally, Hut used AstroSim to explain his research's practical implications in a more educational manner to laypeople. Conducting a typical study in a specialized astrophysics field is difficult because the number of experts is small. So, we consider our observations of these sessions experience reports rather than experiments.

To study double stars' formation, the first session started with a collection of eight stars, representative for a small star-forming region in our galaxy. After a few orbits, simultaneous close encounters between stars cause the formation of both double and triple stars. In most cases, the triple stars are unstable and either fall apart into a single star and a double star or interact through complex exchange processes with other stars. The details of these complex interactions are precisely those that astronomers wish to trace through interactive visualizations.

### *Astronomers Expert in Stellar Dynamics*

One of the first comments our astronomer coauthors made regarding the overall evaluation of AstroSim was that its functionality was comparable to that of the Partiview application. This visualization program by the US National Center for Supercomputing Applications was a response to the desire to project a star cluster on a planetarium dome. In a series of workshops at the Hayden Planetarium at the American Museum of Natural History in New York, the astronomers shared the experience of being immersed in the simulation. They took turns controlling playback and zooming, while discussing the most interesting phenomena they could discern and their theoretical and observational implications.[5] The planetarium visit's largest drawback was the enormous overhead, both in traveling time to and from New York and in the planetarium system setup. Because AstroSim is Second Life-based, in principle a huge number of people anywhere in the world can share the experience without any overhead. They can visit for 10 minutes in between other activities and yet sense the shared copresence of immersion in a virtual 3D online world.

## National Astronomical Observatory of Japan's Visualization Systems

The National Astronomical Observatory of Japan's (NAOJ) Four-Dimensional Digital Universe (4D2U) project is a recent effort to use computer graphics in a dome environment as a tool for public outreach, education, and research. Comprising a set of software applications and hardware devices (domes and theaters), the project has two kinds of applications. The first is the Mitaka space-navigation software, with which users can fly seamlessly from the solar system to the end of the observable universe (see Figure C1). In all the different zoom levels of visualization, Mitaka uses data based on the latest survey projects, such as the Sloane Digital Sky Survey. The second application is computer-simulation animations of astrophysical phenomena, such as forming the universe's large-scale structure, galaxies, individual stars, and planets.

At the Mitaka campus, NAOJ built a dome theater with 3D stereographic-projection capability (http://4d2u.nao.ac.jp/dome), as Figure C2 shows. Both the space navigation software and the simulation animations can run on various hardware, from single-screen notebook PCs to multiscreen stereographic theaters and stereographic full-dome theaters. In Japan, numerous public observatories and science museums have introduced the projection system, which can use the software that 4D2U developed (available for download at http://4d2u.nao.ac.jp/html/program/mitaka/index_E.html).

The 4D2U project has been a great success, particularly as a public-outreach tool. However, the project's original goal was to develop tools for both outreach and research. Unfortunately, the project didn't quite achieve the latter; researchers simply don't use 4D2U's visualization software for their own work. Because the software is rather new and it doesn't provide analysis tools, it might take several years before researchers can use it. Moreover, it's development isn't oriented in that direction because the project's primary goal is to create nice-looking movies.

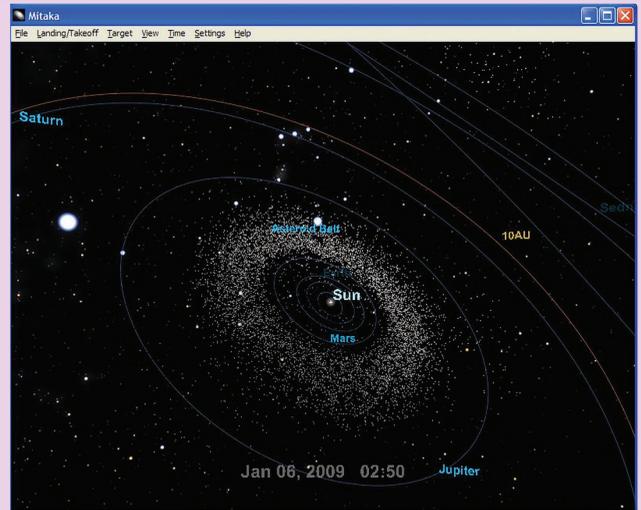

(1)

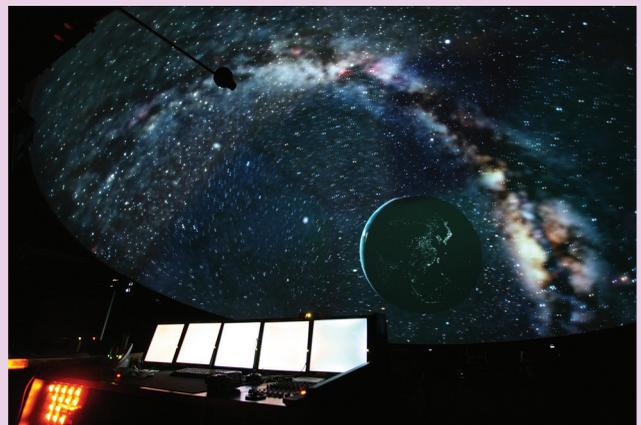

(2)

Figure C. Visualization in a dome environment. (1) The Mitaka software displays a part of the solar system up to the orbit of Saturn. (2) The 4D Digital Universe Dome Theater features a spherical screen and control panel. (Source: National Astronomical Observatory of Japan; used with permission.)

According to the astronomers, a typical way in which astronomers can use systems such as Partiview and NAOJ's Four-Dimensional Digital Universe (4D2U) project to study star clusters is to explore the connections between the microscopic and macroscopic dynamics. (For more on 4D2U, see the "National Astronomical Observatory of Japan's Visualization Systems" sidebar.) Here, the microscopic interactions are the encounters between individual stars, and the macroscopic picture of the stellar dynamics treats the star distribution like a continuum gas. Just as researchers can understand heat conduction in a gas by considering the cumulative effect of the collisions between individual molecules, they can understand the star clusters' overall evolution only as the cumulative effect of all the local encounters. Astronomers can use visualization tools such as Partiview, 4D2U, and AstroSim to inspect microscopic dynamics, which they can then connect with the slow change in the globular star cluster's macroscopic bulk properties. AstroSim's enormous advantage is that astronomers in different parts of the world can perform such an inspection collaboratively—for example, Hut in the US and Makino in Japan.

From the session's start, the stellar-dynamics experts were delighted to be able to move their avatars freely through the simulation, as well as change their camera locations and orientations. They took only a short time to get used to the system, and they quickly learned to start, pause, and restart the system and zoom in and out. The latter feature turned out to be essential for studying how close encounters between three or more stars can crucially change the whole star system's state.

Hut and Makino also appreciated the ability to change time's direction. After an especially interesting encounter, they could replay that encounter backward and forward repeatedly until they charted it in detail, from different directions and distances.

Toward the session's end, a long and interesting discussion started about the degrees of freedom of zooming in and out. Should the centerpoint around which zooming takes place be held fixed in the center of the display area? Would it be better to create a new kind of visible object that users can move and that can serve as a new center for expansion and contraction? Would it be easier to zoom in and out around an avatar's position? Would that be confusing for the other collaborating avatars? This kind of discussion is ongoing. We'll probably implement various choices sometime in the future, given that the astronomers couldn't agree among themselves which options would be most useful.

### *Astronomers Not Expert in Stellar Dynamics*

To assess AstroSim's impact in the astronomy community in general, we organized a session with 10 astronomers from the Meta Institute for Computational Astrophysics (Mica; http://mica-vw.org) who had no stellar-dynamics expertise. In this session, Hut and the astronomers used AstroSim to discuss formation of double and triple stars from a research perspective.

The following are excerpts of the conversation between Hut ("Pema Pera") and an astronomer known in Second Life as "Eamu Godenot" (the terms preceded by `cmd:` represent AstroSim commands):

```
Eamu Godenot: So, can we tell the
difference between forward and back-
ward, once it's in a high-entropy
state?
Eamu Godenot: cmd:play
Pema Pera: oh yes
Pema Pera: binary formation
Pema Pera: only happens forward :)
Eamu Godenot: cmd:pause
Eamu Godenot: Pema, I know that works
over long times....
Pema Pera: if you let it run a bit longer
Eamu Godenot: cmd:play
Pema Pera: you will form some hard
binaries
Eamu Godenot: Bit in these short
snippets?
…
Eamu Godenot: cmd:play
Pema Pera: how about rewinding to see
how the triple formed?
Eamu Godenot: cmd:rewind
Eamu Godenot: cmd:pause
Eamu Godenot: I think it was a four-
body interaction.
Pema Pera: yes, but was it a bound
four-body system?
Pema Pera: or two singles meeting a
binary simultaneously?
Eamu Godenot: I think two singles.
Pema Pera: perhaps rewind a bit further?
Pema Pera: to check?
Eamu Godenot: cmd:rewind
Eamu Godenot: cmd:pause
Pema Pera: complex!!
```

All the astronomers were extremely pleased to see the simulation visualization displaying smoothly. Some of them said they've tried to create the same settings using Second Life's internal scripting methods and had somewhat disappointing results owing to the stars' jerkiness and the overload of computing resources on the Second Life servers. From this conversation that the astronomers had among themselves and the opinions we collected after the session, we conclude that AstroSim was a useful tool for serious research analysis.

### *Individuals Interested in Astronomy*

Finally, to demonstrate AstroSim's capacity for education and outreach, we organized a session with laypeople from the Kira Institute (www.kira.org), an interdisciplinary virtual institute based in Second Life. The Kira Institute unites a diverse group of scholars from many disciplines, including astrophysics, anthropology, psychology, neuroscience, history, literature, art, and philosophy. These scholars are exploring their disciplines' assumptions and limits openly, intellectually, and adventurously.

In this session, Hut explained in simple terms the process of forming double and triple stars in the core of globular star clusters. We had about 30 avatars (users) actively involved not only in discussing the phenomenon, but also, more interestingly, in manipulating the simulation itself (see Figure 6). The following is an excerpt of the discussion between Hut (P) and a participant called Panda Kidd (K):

```
P: the most important commands are:
play, pause, rewind, zoomin, zoomout.
K: cmd:play
P: you can see that initially all stars
```

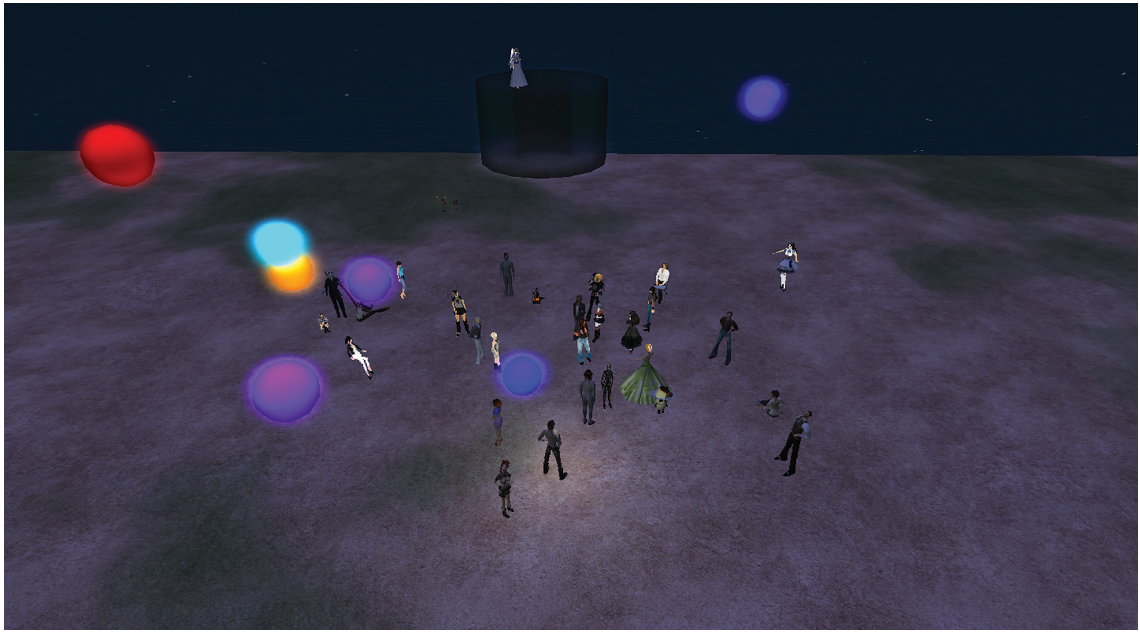

Figure 6. Observing the eight-star simulation in Second Life. Participants (represented as avatars) from the Kira Institute color the stars to appreciate the process of forming a double star.

are single. They sometimes meet each other, but then move on. However, there is a possibility that complex many-body interactions form double stars or even triple stars that remain bound to each other, in tight orbits.
K: ah, I see! There, those two stars just formed a double star. <uses pointing gesture> They keep revolving around each other.
P: yes, I noticed that too. Do you want to check out how this double star actually formed?
K: cmd:pause
K: cmd:rewind
K: okay, this is where they started moving around each other.
P: if you give the two stars a different color, then you can see how they came together.
K: okay. But I then have to go forward in time again, since by now I don't remember which stars had made the double star.
K: cmd:play
K: here they are, forming a bound pair.
K: cmd:pause
K: let me edit the color of the stars. Do I just use standard SL edit commands?
P: yes, just go to edit, then click on textures, and choose from the color palette.
K: okay, I will make them red and green, and the star that interacted with them last I will make yellow.
P: like a traffic light :-)
K: sure, why not.
K: cmd:rewind
K: let me replay the formation of the double star from the time just before the three stars came together.
K: cmd:pause
P: do you know why you need a third star to form a double star?
K: I guess that when two stars meet, they don't stay together?
P: indeed, they would fly away with the same speed that they came in. You need a third star to absorb some of the energy of the two incoming stars, in order to let those two stick together. You will see that the yellow star will shoot out of the interaction region with a higher speed, corresponding to a larger kinetic energy.
K: cmd: play
K: ah, how nice, yes! The red, yellow and green stars come together, dance around for a bit, and then the yellow one moves out, quite fast ...
P: ... leaving the green and red stars bound. You see, you have now witnessed a fundamental process in stellar dynamics: double star formation!

The overall impression we obtained from this heterogeneous group of people is that AstroSim can be an outstanding promotional tool for participatory science as a bridge between astrophysics experts and laypeople.

## Second Life's Limitations

Overall, AstroSim performs simulation visualization smoothly. Nevertheless, owing to Second Life's

IEEE Computer Graphics and Applications 79

operation and data size constraints, our application has some usability and technical limitations.

### Usability Limitations

Annotating stars with textual messages is difficult to implement using virtual-world client technology. Programming skills are required for tasks such as programming scripts to obtain desired behavior in objects. AstroSim's current version doesn't support these more advanced annotation schemes because of the function limitations of libOpenMetaverse, an unofficial, open source Second Life API library. However, color-coding and object description can serve the same purpose.

This version of AstroSim also supports the use of only one set of simulation data files and, therefore, the display of only one simulation phenomenon. Nevertheless, to represent more than one phenomenon for analytical comparison, users can generate one set of data files containing more than one simulation. Specifically, users can generate two or more data frames in a single file, with each frame corresponding to a different simulation. Although this solution reduces flexibility in manipulating the simulations, it's an ideal way to reduce communication resources for Second Life connectivity.

Another usability obstacle, given the virtual space a simulation can occupy, is that avatars can be spread over the entire simulator space, creating problems with using chat channels. However, users can also communicate through instant messaging or voice channels, which chat-distance restrictions don't affect. Using instant messaging rather than voice channels has the advantage of allowing conversation logging, which can be useful when scientists want to keep a record of their discussions.

Finally, techniques common in other visualization applications, such as object interaction, are somewhat limited for scientific visualizations in Second Life because it's primarily a social-networking platform. Fortunately, some visualization techniques, such as camera perspective, visual annotation through coloring, and pointing gestures, are already part of the Second Life viewer technology. Considering that Second Life is an emergent platform for collaborative applications, we'll be able to enhance AstroSim's functionality and usability as developers implement interface improvements in virtual-world client and server software.

### Technical Limitations

We tested our application with a simulation involving 1,024 stars. Although this number is impressive for this type of application, the number of objects is much less than normally expected in a realistic astrophysics simulation. In addition, Second Life restricts the total number of objects, or *prims*, that users can create per island to about 15,000. So, although AstroSim has no hard restrictions in the number of processed objects, this variable constrained us.

Additionally, the size of the XML-based script stream that passed from the data-processing Web service to the visualization-processing Web service was approximately 800 Kbytes per frame. In real simulations, which could involve thousands, if not millions, of stars, the script streams could overflow the network capacity, depending on the connection between both modules. We could have avoided this by merging the two modules, but a core requirement in AstroSim's design was componentization of the solution. This would give us more flexibility in virtual-world connectivity—for example, letting us use other virtual-world platforms, such as OpenSimulator (http://opensimulator.org).

A final limitation is that, from a programming perspective, access to Second Life goes through libOpenMetaverse. Because this library is under active development, some features are either not implemented correctly or not well documented, which limits our ability to provide a solution. For instance, in our first star creation tests, creating 1,000 stars using libOpenMetaverse's object-creation procedure took three to four hours. After refining our algorithm using some low-level commands, we reduced the time to one to two minutes. We expect that with increased stability in the library, we'll be able to provide more-efficient ways to process simulation data.

T o our knowledge, AstroSim is the first generic solution to processing real simulation data for visualization and interaction in a shared 3D environment. By using Second Life, we emphasize the collaboration's participatory aspect—that is, any networked computer user can become a collaborator. No specialized equipment is needed, and registering a Second Life avatar is free.

Although Second Life provided sufficient functionality and stability to implement a working AstroSim prototype, we're closely watching open source initiatives such as OpenSimulator. Such initiatives are expected to overcome Second Life's limitations, especially regarding the number of prims per simulator environment and star-generation processing speed.

We're also continuing to work toward supporting collaboration between research institutes on data-centric applications for environmental science, nanoscience, bioinformatics, and astronomy. Our next step in this area will involve assessing our application's scalability and its adaptation to more intensive data manipulation. For this, we intend to use the Grid infrastructure by the National Research Grid Initiative (Naregi; www.naregi.org/index_e.html). Naregi Grid Middleware version 1.1.1 came out in December 2008.

In addition, we plan to implement advanced navigation and interaction support for collaborators in AstroSim. Envisioned capabilities include efficient wayfinding through landmarks,[6] advanced data selection in 3D space, and natural communication between collaborator avatars. To demonstrate AstroSim's effectiveness, we plan further user tests. We hope that our approach to synchronous collaborative visualization can contribute to making realistic scientific data easily accessible for any interested individual, giving laypeople the opportunity to contribute to science effectively.

You can download a fully working version of our AstroSim prototype at http://www.prendingerlab.net/globallab/?page_id=17.

**Acknowledgments**
*A grant from the Research Organization of Information and Systems, Japan, and a Grand Challenge grant from Japan's National Institute of Informatics partially supported this research. We thank Will Farr from the MIT Kavli Institute for Astrophysics and Space Research and Jeff Ames and Adam Johnson from Genkii for their insightful contributions during this project's requirements phase. Their work in OpenSimulator inspired us to explore using Second Life for similar purposes. Finally, Piet Hut acknowledges receiving a visiting professorship at the National Astronomical Observatory of Japan for 2008.*

**Arturo Nakasone** *is a postdoctoral researcher and project manager at Japan's National Institute of Informatics. His research interests include 3D visualization, virtual storytelling, and virtual-world and game interactivity. Nakasone has a PhD in information science from the University of Tokyo. Contact him at arturonakasone@nii.ac.jp.*

**Helmut Prendinger** *is an associate professor at Japan's National Institute of Informatics. His research interests include human-computer interaction, 3D Internet, and multimodal-content creation in virtual worlds. Prendinger has a PhD in the logic and philosophy of science and artificial intelligence from the University of Salzburg. Contact him at helmut@nii.ac.jp.*

**Simon Holland** *is an internship student at Japan's National Institute of Informatics. His research interests include Second Life, 3D visualization applications, and computer graphics. Holland has a master's in informatics from the University of Edinburgh. Contact him at simon.holland@gmail.com.*

**Piet Hut** *is a professor of astrophysics and interdisciplinary studies at the Institute for Advanced Study in Princeton, New Jersey. His research interests include large-scale simulations in stellar dynamics, as well as collaborative interdisciplinary research from geology and paleontology to cognitive science and philosophy. Hut has a PhD in astrophysics from the University of Amsterdam. Contact him at piet@ias.edu.*

**Jun Makino** *is a professor at the National Astronomical Observatory of Japan and director of Japan's Center for Computational Astrophysics. His research interests include computational astrophysics and computer architecture. Makino has a PhD in astrophysics from the University of Tokyo. Contact him at makino@cfca.jp.*

**Kenichi Miura** *is a professor of high-end computing at Japan's National Institute of Informatics and project leader of the Resources Linkage for e-Science project. His research interests include grid technology and supercomputing and special-purpose processor architecture development. Miura has a PhD in computer science from the University of Illinois at Urbana-Champaign. Contact him at kenmiura@grid.nii.ac.jp.*